\documentclass[fleqn,twoside,twocolumn,nofootinbib,showkeys]{revtex4} 
\usepackage[nocpr]{ujp} 
\begin{document}
\title[Parametric Excitation of the Surface Magnetostatic Modes]
{PARAMETRIC EXCITATION\\ OF SURFACE MAGNETOSTATIC MODES\\ IN AN AXIALLY MAGNETIZED ELLIPTIC\\ CYLINDER UNDER LONGITUDINAL PUMPING}%
\author{M.A.~Popov}
\affiliation{Taras Shevchenko National University of Kyiv}
\address{64/13, Volodymyrs'ka Str., Kyiv 01601, Ukraine}
\email{maxim_popov@univ.kiev.ua}

\udk{537.635}  \razd{\secvii}

\autorcol{M.A.~Popov}%

\setcounter{page}{452}%

\begin{abstract}
A rigorous analytical theory of parametric excitation under the
longitudinal pumping has been developed for the surface
magnetostatic modes of a long elliptic ferrite cylinder magnetized
along its axis with regard for the boundary conditions at the
surface of the cylinder.\,\,It is shown that a pair of
frequency-degenerated counter-propagating surface modes at half the
pumping frequency can be parametrically excited, and the expressions
for the corresponding parametric excitation threshold have been
derived.\,\,The threshold demonstrates a strong dependence on the
mode number and elliptic cylinder's aspect ratio and tends from
above for the large aspect ratio to the value deduced on the basis
of the plane-wave analysis.\,\,The simple analytical relation
between the ratio of axes of the high-frequency magnetization
polarization ellipse of excited surface magnetostatic oscillations
and the parametric excitation threshold is obtained, discussed, and
graphically illustrated.
\end{abstract}


\maketitle

\section{Introduction}\vspace*{1mm}
The parametric excitation of spin waves due to the intrinsic
nonlinear properties of ferromagnetic materials plays a key role in
practical applications.\,\,While the magnetization oscillations with
small amplitudes can be safely analyzed in the linear approximation
\cite{1}, the nonlinear properties of ferrite for relatively large
amplitudes of the high-frequency magnetization lead to various
nonlinear effects \cite{2}, including the parametric excitation (PE)
of spin waves \cite{3}.\,\,On the one hand, such phenomenon
restricts the dynamic range of the input RF power of magnetostatic
resonators.\,\,On the other hand, a number of nonlinear devices,
such as a power limiter and a signal-to-noise enhancer are based on
this effect \cite{4}.\,\,Therefore, the careful examination of
parametric excitation with regard for the specific features of a
ferrite resonator and excitation conditions is of importance for the
applied \mbox{research.}\looseness=1

Suhl \cite{3} developed the basics of the PE theory, as applied to a
transversely pumped isotropic ferromagnet.\,\,Subsequently, the
theoretical model was improved to account for the arbitrary
orientation and polarization of a microwave pumping field
\cite{5}.\,\,Finally, it was generalized for the first and second
bands under the dual pumping of ferrite materials with either
uniaxial or cubic magnetocrystalline anisotropy, by using arbitrary
polarized and oriented RF fields~\cite{6}.

However, Suhl's theory utilizes the expansion of the high-frequency
magnetization in uniform plane spin waves, which is justified only
when the wave number $k$ of excited spin waves is much larger than
the inverse dimensions of a sample.\,\,But, for thin ferrimagnetic
films with thickness of the order of a few to a few tens of microns,
the typical wave numbers of resonator eigen-excitations~-- surface
magnetostatic oscillations (SMSO)~-- are much less than the inverse
thickness.\,\,In this case, one ought to expand the magnetization
vector ${\bf m} $ and the RF magnetic field in problem's normal
modes  ${\bf m}_n$: ${\bf m} = \sum_n {( {A_n {\bf m}_n  + {\rm
c.c.}} )} $ \cite{7,8}, instead of plane spin waves.

In this paper, the parametric excitation of SMSO in a longitudinally
magnetized yttrium-iron garnet (YIG) film resonator with elliptic
cross-section under the longitudinal pumping will be considered, by
taking the actual boundary conditions at the resonator surface into
account.\,\,The ferrite anisotropy is neglected, by assuming the
external static magnetic field to be much larger than typical YIG
cubic and uniaxial anisotropy fields ($\approx$50 Oe).

\section{General Theory}

The exact analytical theory of SMSO in infinitely long isotropic ferrite resonator
magnetized along its axis with elliptic
cross-section (Fig.~1) in the nonexchange limit is presented in
\cite{9}.

It was shown \cite{9,10}, that the eigenfrequency of the SMSO $n$-th
mode of such resonator is given by:
\begin{equation}
\omega _n^2  = (\omega _H  + \omega _M /2)^2  - 1/4\omega _M^2 ( {(a
- b)/(a + b)} )^{2|n|}\!.
\end{equation}

The magnetostatic modes of the infinitely long elliptic resonator
can be characterized by three indices \cite{7}, namely, the number
of nodes in the circumferential direction $n$, index $r$ of a
solution of the characteristic equation, and wavenumber $ \beta $
corresponding to the propagation along the cylinder axis.\,\,For the
axially uniform oscillations, $ \beta =0,$ and the surface modes are
labeled by $r=0,$ according to \cite{7}.\,\,Hereafter, we will
designate each mode with the single subscript $n$ instead of all
three indices $(n,0,0)$, for the sake of brevity.

Since $ \omega _n  = \omega _{-n}$ (1), the external pumping RF
magnetic field ${\bf h}$ (applied in parallel to the DC field ${\bf
H}_0$) with the frequency $\omega _p  = 2\omega _n $ can
parametrically excite two frequency-degenerated counterpropagating
surface magnetostatic modes with the indices $n$ and~$-n$.

The magnetostatic potential $\Psi $ for the SMSO $n$-th mode in an elliptic
ferrite cylinder with semiaxes $a$ and $b$ can be expressed in the
modified elliptic coordinate system $(\rho ,\phi ,z)$ as \cite{9}
\[
\Psi _n (\rho ,\phi ) = B_n(R_n^ +  (\rho )\cos (n\phi ) \,-
\]\vspace*{-7mm}
\begin{equation}
 -\,i{\mathop{\rm sgn}} (n)\frac{{\mu  + 1/A}}{{\mu _a }}R_n^ - (\rho
)\sin (n\phi ) ),
\end{equation}
where $B_n$ is the mode amplitude, $R_n^ +  (\rho ) = (\rho ^{|n|} +
{+\,( {c/2} )^{2|n|} \rho ^{ - |n|} )}$, $R_n^ -  (\rho ) = (\rho
^{|n|} - ( {c/2} )^{2|n|} \rho ^{ - |n|} )$, $A = ( {1 - ( {c/(a +
b)} )^{2|n|} } ) / ( {1 + ( {c/(a + b)} )^{2|n|} } $, $c =$ $= \sqrt
{a^2  - b^2 } $, $\mu  = (\omega ^2 - \omega _{^1 }^2 )/(\omega ^2 -
\omega _{H }^2 )$, $\mu _a  = {= { \omega \omega _{M } /(\omega^2 -
\omega _{H }^2 )}}$, $\omega _{1}^2 = \omega _H (\omega _H  + \omega
_M )$, $\omega _M = {= \gamma 4\pi M_0 }$, $\omega _H  = \gamma H_0
$, $\gamma$ is the gyromagnetic ratio, and $M_0$ is the saturation
magnetization.\,\,In (2), one should treat $ \omega $ as the
eigenfrequency $\omega_n$ of the $n$-th mode at a given magnetic
field $H_0$ (see (1)).

\begin{figure}
\vskip1mm
\includegraphics[width=6cm]{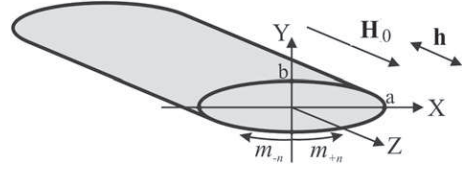} 
\vskip-3mm\caption{Longitudinally magnetized ferrite elliptic
cylinder under parallel pumping}
\end{figure}

In view of the standard expressions
\[
{\bf h}_n={\rm grad}\, \Psi _n ,\quad {\bf m}_n=\hat \chi {\bf h}_n
, \]\vspace*{-7mm}
\[
\hat \chi= \displaystyle{\frac{{\omega _M }}{{\omega _H^2 - \omega
^2 }}} \left(\!\! {\begin{array}{ll}
   {\omega _H } & {i\omega }  \\
   { - i\omega } & {\omega _H }  \\
\end{array}} \!\!\right)\!\!
,
\]
relations (2) yield the explicit formulae for the components of the
high-frequency magnetization vector ${\bf m}_n$ for each SMSO mode.
Then we verified,by straight\-for\-ward calculations,  the
orthogonality relation $\int {\left( {m_{\rho n} m_{\varphi m}-
m_{\varphi n} m_{\rho m} } \right)dV}=0$ and calculated the
eigenmode normalization constant $D_n=$ $=- i\int ( m_{\rho n}
m_{\varphi n}^* - m_{\rho n}^* m_{\varphi n}  )dV $ \cite{7}.\,\,It
was found, that $D_n=D_{ - n}={\frac{{2\pi C_n^2 }}{{|n|}}}E_n $,
where $C_n $ is some expression depending on the frequency, magnetic
field, saturation magnetization, and geometric parameters of a
sample, and $E_n = 2\omega _n  ( 2\omega _H  + \omega _M ( 1 + ( (a
- b)/(a \,+$ $+\, b) )^{|n|}  ) )^{ - 1}\! $.\,\,Note that, in
modified elliptical cylindrical coordinates \cite{9}, we have $dV
\equiv \rho h_\rho ^2 d\rho d\varphi dz$, $h_\rho =$ $= \sqrt {( {1
- {c^2} /{4\rho ^2 } } )^2 + ({c^2} / {4\rho ^2})  \sin ^2 \varphi
}$.

In \cite{7}, the general expression for the experimentally observed
parametric excitation threshold RF field $h_c$ was found to be
 \begin{equation}
 \left( {\gamma h_c } \right)^2  = \frac{{4\omega _{rn}\omega _{rm}}}{{\lambda _{n,m} \lambda _{m,n}^*
 }},
\end{equation}
where $\lambda _{n,m}  = (1/D_n )\int {\left( {{\bf m}_n^* {\bf
m}_m^* } \right)dV}$, and $\omega_{rn}$ is the relaxation frequency of the
proper mode.

Using (2), we obtain
\[
\int \left( {m_{\rho n}^* m_{\rho m}^*  + m_{\varphi n}^* m_{\varphi
m}^* } \right)dV = \frac{{\pi C_n^2 }}{{|n|}}(1 - E_n^2 ),
\]
when $|n|=|m|$, and is equal to zero otherwise (here, $C_n$ and
$E_n$ are exactly the same as in the expression for
$D_n$).\,\,Therefore,
\[
\lambda _{n,m}  = \frac{{(1 - E_n^2 )}}{{2E_n }}\delta _{|n|,|m|}  =
\lambda _{m,n}^*.
\]

This means that the PE process like $\omega _p=\omega _n+\omega _m$,
$|n| \ne |m|$, which could be allowed by the energy conservation
law, would have, nevertheless, the infinite threshold due to the
zero overlapping integral $\lambda _{n,m}$.\,\,Thus, only the
parametric excitation of two SMSO modes having opposite azimuthal
indices $n=-m$ is allowed and would be considered further. By
substituting all the previously calculated expressions into (3), we
obtain
\[ h_c  = \left(\! {\frac{{2\omega _{rn}^{} }}{\gamma
}} \!\right)\frac{{2E_n }}{{(1 - E_n^2 )}}.
\]

After some cumbersome calculations, the final expression takes the form
\begin{equation}
h_c=\left(\! \frac{{2\omega_{rn}}}{{\gamma}}\!\right)\frac{{2\omega
_n }}{{\omega _M \left(\! \displaystyle{\frac{{a - b}}{{a + b}}}
\!\right)^{\!\!|n|} }} =   \frac{{\eta \omega _p }}{{\omega _M
\left(\! \displaystyle{\frac{{a - b}}{{a + b}}} \!\right)^{\!\!|n|}
}}\Delta {H}_k,
\end{equation}
where $\eta  = \partial \omega _n /\partial \omega _H  = (\omega _H +
\omega _M /2)/\omega _n $ is the ellipticity factor \cite{11},
and $\Delta {H}_k$ is the ferromagnetic resonance linewidth.

Apparently, the excitation threshold for SMSO under the longitudinal
pumping strongly depends on the geometric parameters of a sample
(e.g., the aspect ratio $a/b$).\,\,But otherwise, expression (4) is
similar to that deduced on the basis of the plane-wave
\mbox{analysis}~\cite{1}.

As it was pointed out earlier \cite{8}, the PE process efficiency
strongly correlates with the polarization of excited spin
waves.\,\,Next, we will elucidate this statement for the problem
under consideration and express it in the strict mathematical form.

Using the explicit expressions for ${\bf m}_n $, our calculations show
that the ratio of axes of the high-frequency magnetization polarization
ellipse is given by the formula
\[
m_{\min } /m_{\max }  = \tan (1/2\arcsin (2E_n /(1 + E_n^2 ))).
\]

It is worth noting that $m_{\min } /m_{\max }$ does not depend on
coordinates, i.e., it is spatially uniform.

After some simplifications, we obtain a formula that explicitly
expresses the ratio of axes of the eigenmode polarization ellipse in terms of
the magnetic and geometric parameters of the SMSO resonator:
\begin{equation}
\frac{{m_{\min}}}{{m_{\max} }}=\tan\left(\!{1/2\arcsin
{\frac{{\omega_n}}{{\omega_H+ {\frac{\omega_M}{2}}}}} }
\!\right)\!\!.
\end{equation}

Considering the expression for the ellipticity factor and relation (1), one can
see that magnetization's polarization is defined by the ellipticity
factor $\eta$ only, according to  $m_{\min } /m_{\max }  = \tan
\left( {1/2\arcsin \left( {1/\eta } \right)} \right)$.

Since the critical field $h_c$ and the polarization state
depend on the same coefficient $E_n$, we can express one physical
quantity directly via another one. Thus, we have
\begin{equation}
h_c  = \left(\! {\frac{{2\omega _{rn}}}{\gamma }} \!\right)
\frac{{m_{\min} /m_{\max} }}{{1 - \left(m_{\min} /m_{\max} \right)^2
}}.
\end{equation}

The physical origin of such correlation is clear: for the circular
precession of the magnetization ($m_{\min } /m_{\max }  = 1$), the
longitudinal component of the magnetization $m_z$  is absent, and no
coupling with the pumping field is possible ($h_c  \to \infty $).
For a more elliptic precession, $m_z$  becomes correspondently
larger.\,\,Hence, the interaction is stronger, and the threshold is
lower~\cite{8}.

\section{Discussion}

Let us consider two limiting cases of (4): a circular ferrite rod
($a=b$) and a very elongated elliptic cylinder ($a\gg b$).\,\,In the
first case, $h_c  \to \infty $, since the characteristic equation
admits a solution only for $n>0$, and a pair of counterpropagating
surface modes required for the PE process is absent, as it was
correctly pointed out in \cite{7}.\,\,As for the second case, let us
use the previously published expression for the PE of traveling
surface magnetostatic waves with the wavevectors $ \pm k$ in a thin
magnetic film with thickness $d$ \cite{12}.\,\,In that situation,
the threshold is equal to ${h_c = \left( {2\omega _{rn}^{} /\gamma }
\right)\left( {\omega _p /\omega _M } \right)\exp (|k|d)}$, which
for ${kd\ll1}$ reduces to ${h_c = \left( {2\omega _{rn}^{} /\gamma }
\right)\left( {\omega _p /\omega _M } \right)(1+|k|/d)}$.\,\,On the
other hand, expression (4) for ${b/a\ll1}$ reduces to ${h_c = \left(
{2\omega _{rn}^{} /\gamma } \right)\left( {\omega _p /\omega _M }
\right)(1+2b|n|/a)}$.\,\,Those two formulae would be identical, if
we make the natural replacement $2b \to d$ and assume that an
``equivalent'' wavevector $|k|=|n|/a $ can be assigned to each
eigenmode with index $n$.\,\,Since $1/a \to 0,$ the discrete set of
mode indices smoothly transforms into a continuous manifold of $k$.
Thus, expression (4) gives the correct results in both limiting
cases.

The dependence of the threshold on the cylinder shape is illustrated
in Fig.~2, where the normalized microwave threshold field $h_c
/\Delta H_k $ for a few lowest-order SMSO modes is depicted as a
function of the aspect ratio $a/b$.\,\,In calculations, we used
$H_0=1$~kOe and the value of $4\pi M_0  =1.75$~kG typical of YIG.
The dash-dotted line shows the normalized threshold for infinite
isotropic media (that is equal to $\omega _p /\omega _M $), by
assuming that $\Delta H_k $ in both cases are identical.\,\,One can
see that the parametric excitation threshold drastically increases
for cylinder's shape close to the circular one.\,\,But, for elliptic
cylinders with a large aspect ratio (for example, thin-film
resonators), it is approaching the value for plane spin waves.
Spe\-ci\-fi\-cal\-ly, for the $n=1$ mode, the difference from the
``bulk'' value is less than 15\% for $a/b>20$.\,\,Moreover, the
threshold noticeably increases with the mode
\mbox{number.}\looseness=1

Thus, the calculations presented here allow one to evaluate the PE
threshold for any given mode of an elliptic resonator and give the
more flexibility to an SMSO resonator designer in choosing the
dynamic power range of a device.\,\,For example, if the operation at
a larger input power is required, the resonator, according to (4)
and Fig.\,\,2, should work on higher modes with large $h_c$ or must
be manufactured as a circular cylinder.\,\,On the other hand, for
the applications like a power limiter, one can precisely set the
desired resonator's threshold power, by simply selecting the
appropriate axis ratio (see Fig.\,\,2).

The analysis of expression (5) demonstrates that, for a circular
cylinder ($a=b$), all modes without exception have circular
polarization.\,\,However, when cylinder's aspect ratio $a/b$
increases, the $m_{\min }/m_{\max }$ ratio start decreasing, and the
modes with larger index $n$ are always being more
``circular''.\,\,In addition, the polarization ellipse aspect ratio
increases with $H_0$, tending to 1 for the large bias (see
Fig.\,\,3).

The threshold vs.\,\,polarization dependence, as described by
expression (6), is illustrated in Fig.\,\,4.\,\,It is clearly seen
that the more elliptic precession of the magnetization (smaller
$m_{\min } /m_{\max }$) facilitates, indeed, PE under the
longitudinal pumping, as was pointed out earlier.\,\,Note that the
very elliptic (close to linear) precession is beyond the scope of
the current theory, since  the assumption  $m_z \ll m_x,m_y $ ($m_i$
being the dynamical (high-frequency) components of the
magnetization) used when deriving the expressions for tensor
magnetic permeability is no longer valid in this case.

\begin{figure}
\vskip1mm
\includegraphics[width=7.5cm]{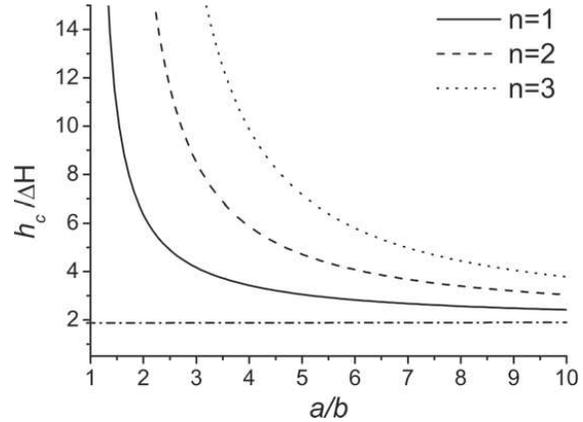} 
\vskip-3mm\caption{Normalized microwave threshold field as a
function of the elliptic cylinder aspect ratio}
\end{figure}
\begin{figure}
\vskip4mm
\includegraphics[width=7.5cm]{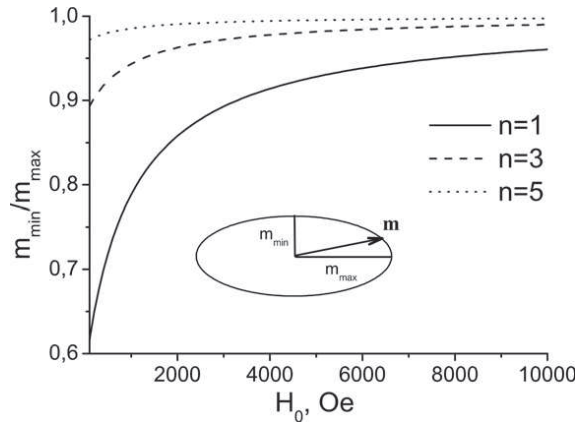} 
\vskip-3mm\caption{Modes magnetization ellipse axis ratio as a
function of the bias magnetic field (cylinder's aspect ratio
$a/b=3$)}
\end{figure}

\begin{figure}[h!]
\vskip4mm
\includegraphics[width=7.5cm]{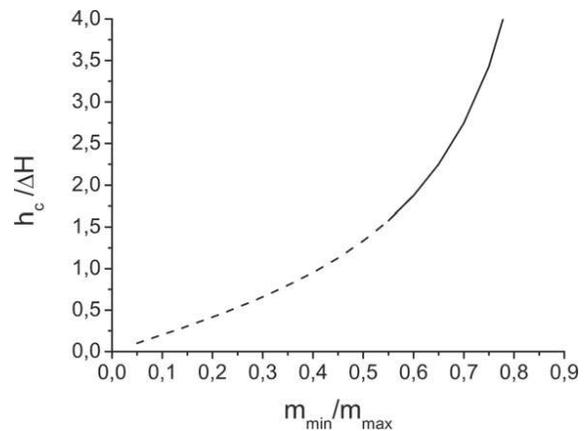} 
\vskip-3mm\caption{Parametric excitation threshold vs.\,\,the
polarization axis ratio of excited magnetostatic
oscillations}\vspace*{-6mm}
\end{figure}

In order to define the limits of current theory's applicability,
the investigation of the longitudinal and transversal high-frequency
components of the magnetization vector, assuming $| {{\bf M}} | =
$~const, was done, resulting in the following formula:
\begin{equation}
\frac{{m_{\min } }}{{m_{\max } }} = \sqrt {4\left(\! {\frac{\alpha
}{{m_{x0} }}} \!\right)^{\!\!2}  + 1}  - 2\frac{\alpha }{{m_{x0} }}.
\end{equation}

Expression (7) defines the polarization ellipse aspect ratio $
m_{\min } /m_{\max },$ for which $m_z$ reaches a value equal to
${\alpha\, m_y}$,  $0<\alpha<1$, for the given polarization ellipse
normalized major semiaxis $m_{x0}=m_x /M_0$. The parameter $\alpha=
m_z/m_y$ determines the smallness of the longitudinal component
$m_z$ relative to the transversal component $m_y$.\,\,If
$\alpha\ll1$, the standard expressions for tensor magnetic
permeability are entirely valid.\,\,Otherwise, those expressions are
no longer applicable, and all the theoretical results presented here
are doubtful.\,\,Formula (7) allows one to estimate the range of $
m_{\min } /m_{\max }$, for which our theoretical model remains
correct.\,\,For example, for fixed $m_x/M_0=0.1$, $\alpha$ is equal
$0.05$ for $m_{\min}/m_{\max}=0.41,$ and $\alpha=0.1$ for $m_{\min}
/m_{\max}=0.24.$ Thus, the safe interval is roughly $0.5<m_{\min}
/m_{\max}<1$.\,\,For smaller $m_x/M_0$ we will always get lesser
values of the lower boundary of $m_{\min}/m_{\max}$.\,\,Therefore,
it would be safe to assume that, for relatively small  $m_x /M_0$
(which is typical of the parametric excitation processes under the
parallel pumping), the curve in Fig.\,\,4 is trustworthy for
$m_{\min}/m_{\max}$ above approximately~0.5.\looseness=1

\section{Conclusions}

Analytical calculations by the theory of parametric excitation of
magnetostatic surface oscillations in longitudinally magnetized
elliptic cylinders under the longitudinal pumping have been
conducted.\,\,The final expressions are obtained in the simple
convenient form suitable for the further analysis.

The parametric excitation threshold for various mode numbers and
cylinder aspect ratios has been derived and analyzed.\,\,It is shown
that the parametric excitation threshold for SMSO in a thin ferrite
film is of the same order of magnitude with that calculated within
the classical theory for plane spin waves (SW).\,\,The
interpretation of the experimental results and the thorough analysis
of both possible mechanisms of parametric excitation are carried
out.\,\,In\-deed, we have the relation $h_c^{\rm SMSO} /h_c^{\rm SW}
= $ $={\left(\! {\Delta {H}_k^{\rm SMSO} /\Delta {H}_k^{\rm SW} }
\!\right)\! {\left( {a + b} \right)^{|n|} } / {\left( {a - b}
\right)^{|n|} } }$.\,\,For example, for the pumping field frequency
$\omega_p=10$~GHz, the parametric SMSO excited at $\omega_p /2$ in
an YIG resonator biased with $H_0=1000$~Oe will have low $k$ and
$\Delta {H}_{k \to 0}^{\rm SMSO}\approx0.25$~Oe \cite{13}.\,\,At the
same time, the plane spin waves with equal frequency would have
$k\approx10^5$~cm$^{-1}$ and much larger $\Delta {\rm H}_k^{\rm
SW}\approx 0.7$~Oe due to the additional contribution from the
dipolar 3-mag\-non confluence process \cite{14}.\,\,In this
situation, the SMSO main mode ($n=1$) in a resonator with aspect
ratio $a/b>2$ will have a lower parametric excitation threshold than
plane spin
\mbox{waves.}

The analytical expression for the ratio of axes of the
high-frequency magnetization polarization ellipse is obtained, and
the correspondence between the polarization state and the PE
threshold is investigated.\,\,The expression directly connecting the
ratio of axes and the PE threshold is found and graphically
illustrated, and the bounds of its applicability are indicated.

Earlier \cite{9}, it was shown that it the nonexchange limit SMSO
spectrum of a long YIG longitudinally biased resonator with
rectangular cross-section can be calculated with the use of the
geometric approximation of the resonator cross-section with
inscribed ellipse.\,\,Thus, the presented theory, though being
derived for an elliptic resonator, can be potentially applied to the
widely used film ferrite resonators with rectangular shape.

\vspace*{-2mm}
\rezume{
М.О.\,Попов}{ПАРАМЕТРИЧНЕ ЗБУДЖЕННЯ\\ ПАРАЛЕЛЬНОЮ НАКАЧКОЮ
ПОВЕРХНЕВИХ\\ МАГНІТОСТАТИЧНИХ КОЛИВАНЬ\\ В ПОЗДОВЖНЬО
НАМАГНІЧЕНОМУ\\ ЕЛІПТИЧНОМУ ЦИЛІНДРІ} {Розроблено аналітичну теорію
параметричного збудження паралельною накачкою поверхневих
магнітостатичних коливань нескінченно довгого еліптичного
феромагнітного циліндра, намагніченого вздовж осі, з урахуванням
граничних умов на поверхні феромагнетика. Показано можливість
параметричного збудження пари вироджених мод з протилежними
напрямками поширення, з частотами, що дорівнюють половині частоти
накачки, та отримано вирази для порога цього процесу. Знайдено, що
порогова амплітуда поля накачки сильно залежить від номера моди та
відношення великої та малої півосі еліптичного циліндра і при
великому значенні цього відношення прямує зверху до величини, що
розрахована на основі моделі плоских хвиль. Було отримано,
проаналізовано та графічно проілюстровано просте аналітичне
співвідношення між еліптичністю поляризації високочастотної
намагніченості збуджених поверхневих магнітостатичних коливань та
порогом їх параметричного збудження. }

\end{document}